\documentclass[12pt]{article}
\usepackage{natbib}
\usepackage{graphicx}
\usepackage{amssymb,amsfonts,amsmath}
\usepackage{geometry}
\usepackage{setspace}

\usepackage{multirow}
\begin{document}
\title{Extended Poisson-Tweedie: properties and regression models for count data}
\author{Wagner H. Bonat\thanks{Department of Mathematics and Computer Science, University of Southern Denmark, Odense, Denmark.
		    Department of Statistics, Paran\'a Federal University, 
		    Curitiba, Paran\'a, Brazil. E-mail: wbonat@ufpr.br} 
~and Bent J{\o}rgensen\thanks{Department of Mathematics and Computer Science, University of Southern Denmark, Odense, Denmark.}
~and C\'elestin C. Kokonendji\thanks{Universit\'e de Franche-Comt\'e, Laboratoire de Math\'ematiques de Besan\c{c}on, 
		    Besan\c{c}on, France.} \\
~and John Hinde\thanks{National University of Ireland, School of Mathematics, Galway, Ireland.}
~and Clarice G. B. Dem\'etrio\thanks{S\~ao Paulo University, S\~ao Paulo, Brazil.}}

\date{\vspace{-5ex}}
\maketitle

\begin{abstract}
We propose a new class of discrete generalized linear models based on the class of 
Poisson-Tweedie factorial dispersion models with variance of the form $\mu + \phi\mu^p$, 
where $\mu$ is the mean, $\phi$ and $p$ are the dispersion and Tweedie power parameters, 
respectively. The models are fitted by using an estimating function approach obtained
by combining the quasi-score and Pearson estimating functions for estimation of the
regression and dispersion parameters, respectively. This provides a flexible and efficient
regression methodology for a comprehensive family of count models including Hermite, 
Neyman Type A, P\'olya-Aeppli, negative binomial and Poisson-inverse Gaussian.
The estimating function approach allows us to extend the Poisson-Tweedie distributions to 
deal with underdispersed count data by allowing negative values for the dispersion 
parameter $\phi$. Furthermore, the Poisson-Tweedie family can 
automatically adapt to highly skewed count data with excessive zeros, 
without the need to introduce zero-inflated or hurdle components, by the simple estimation 
of the power parameter. Thus, the proposed models offer a unified framework to deal with
under, equi, overdispersed, zero-inflated and heavy-tailed count data.
The computational implementation of the proposed models is fast, relying only on a simple Newton 
scoring algorithm. Simulation studies showed that the estimating function approach provides
unbiased and consistent estimators for both regression and dispersion parameters.
We highlight the ability of the Poisson-Tweedie distributions to deal with count data 
through a consideration of dispersion, zero-inflated and heavy tail indexes, and illustrate its
application with four data analyses.
We provide an \texttt{R} implementation and the data sets as supplementary materials.
\end{abstract}

\section{Introduction}
Generalized linear models (GLMs)~\citep{Nelder:1972} have been the main 
statistical tool for regression modelling of normal and non-normal data 
over the past four decades. The success enjoyed by the GLM framework
comes from its ability to deal with a wide range of normal and non-normal 
data. GLMs are fitted by a simple and efficient Newton score algorithm relying
only on second-moment assumptions for estimation and inference.
Furthermore, the theoretical background for GLMs is well established in the class
of dispersion models~\citep{Jorgensen:1987, Jorgensen:1997} as a generalization 
of the exponential family of distributions. 
In particular, the Tweedie family of distributions plays an important role in the 
context of GLMs, since it encompasses many special cases including the normal, 
Poisson, non-central gamma, gamma and inverse Gaussian. 

In spite of the flexibility of the Tweedie family,
the Poisson distribution is the only choice for the analysis
of count data in the context of GLMs. For this reason, in practice there is probably an 
over-emphasis  on the use of the Poisson distribution for count data. 
A well known limitation of the Poisson distribution is its mean and variance 
relationship, which implies that the variance equals the mean, referred to as equidispersion.
In practice, however, count data can present other features, namely 
underdispersion (mean $>$ variance) and overdispersion (mean $<$ variance) that is 
often related to zero-inflation or a heavy tail. 
These departures can make the Poisson distribution unsuitable, or at least of limited use, for 
the analysis of count data. The use of the Poisson distribution for non-equidispersed
data may cause problems, because, in case of overdispersion, standard errors calculated under 
the Poisson assumption are too optimistic and associated hypothesis tests will tend to give false 
positive results by incorrectly rejecting null hypotheses. 
The opposite situation will appear in case of underdispersed data. 
In both cases, the Poisson model provides unreliable standard
errors for the regression coefficients and hence potentially misleading inferences.

The analysis of overdispersed count data has received much attention.
\citet{Hinde:1998} discussed models and estimation algorithms for overdispersed data.
\citet{Kokonendji:2004,Kokonendji:2007} discussed the theoretical aspects of 
some discrete exponential models, in particular, the Hinde-D\'emetrio and Poisson-Tweedie classes. 
\citet{Shaarawi:2011} applied the Poisson-Tweedie family for modelling species abundance.
\citet{Rigby:2008} presented a general framework for modelling overdispersed count data,
including the Poisson-shifted generalized inverse Gaussian distribution. 
\citet{Rigby:2008} also characterized many well known distributions, such as the negative
binomial, Poisson-inverse Gaussian, Sichel, Delaporte and Poisson-Tweedie as Poisson
mixtures. In general, these models are computationally slow to fit to large data sets,
their probability mass functions cannot be expressed explicitly and they deal only with 
overdispersed count data. Further approaches include the normalized tempered stable 
distribution~\citep{Kolossiatis:2011} and the tempered discrete Linnik 
distribution~\citep{Barabesi:2016}.

The phenomenon of overdispersion is in general manifested through a heavy tail and/or zero-inflation.
\citet{Zhu:2009} discussed the analysis of heavy-tailed count data based on the 
Generalized Poisson-inverse Gaussian family.
The problem of zero-inflation has been well discussed~\citep{Ridout:1998} and solved by including hurdle
or zero-inflation components~\citep{Zeileis:2008}. These models are specified by
two parts. The first part is a binary model for the dichotomous event of having zero 
or count values, for which the logistic model is a frequent choice. 
Conditional on a count value, the second part assumes a discrete distribution, 
such as the Poisson or negative binomial~\citep{Loeys:2012}, or zero-truncated versions for the hurdle model.
While quite flexible, the two-part approach has the disadvantage of
increasing the model complexity by having an additional linear predictor to describe
the excess of zeros. 

The phenomenon of underdispersion seems less frequent in practical data analysis,
however, recently some authors have given attention towards the underdispersion
phenomenon. \citet{Sellers:2010} presented a flexible regression model based on
the COM-Poisson distribution that can deal with over and underdispersed data.
The COM-Poisson model has also recently been extended to deal with zero-inflation~\citep{Sellers:2016}.
\citet{Zeviani:2014} discussed the analysis of underdispersed experimental data 
based on the Gamma-Count distribution. Similarly, \citet{Kalktawi:2015} proposed a discrete 
Weibull regression model to deal with under and overdispersed count data.
Although flexible, these approaches share the disadvantage that the probability
mass function cannot be expressed explicitly, which implies that estimation and inference
based on the likelihood function is difficult and time consuming. Furthermore, the expectation
is not known in closed-form, which makes these distributions unsuitable for regression modelling,
where in general, we are interested in modelling the effects of covariates on a function of the 
expectation of the response variable.

Given the plethora of available approaches to deal with count data  in the literature,
it is difficult to decide, with conviction, which is the best approach for a particular
data set. The orthodox approach seems to be to take a small set of models, such as the
Poisson, negative binomial, Poisson-inverse Gaussian, zero-inflated Poisson, zero-inflated
negative binomial, etc,  fit all of these models and then choose the best fit by using
some measures of goodness-of-fit, such as the \textit{Akaike} or Bayesian information criteria.
A typical example of this approach can be found in~\citet{Oliveira:2015},
where the authors compared the fit of eight different models for the analysis of
data sets related to ionizing radiation. Although reasonable, such an approach is difficult
to use in practical data analysis. The first problem is to define the set of models to be
considered. Second, each count model can require specific fitting algorithms and give its own set
of fitting problems, in general due to bad behaviour of the likelihood function.
Third, the choice of the best fit may not be obvious, with different information criteria leading
to different selected models. Finally, the uncertainty around the choice of distribution is not taken into account
when choosing the best fit. Thus, we claim that it is very useful and attractive to have a unified model 
that can automatically adapt to the underlying dispersion and that can be easily implemented 
in practice.

The main goal of this article is to propose such a new class of count generalized linear 
models based on the class of Poisson-Tweedie factorial dispersion models~\citep{Jorgensen:2016} with 
variance of the form $\mu + \phi\mu^p$, where $\mu$ is the mean, $\phi$ and $p$ are 
the dispersion and Tweedie power parameters, respectively. 
The proposed class provides a unified framework to deal with over-, equi-, or underdispersed, 
zero-inflated, and heavy-tailed count data, with many potential applications.

As for GLMs, this new class relies only on second-moment assumptions 
for estimation and inference. The models are fitted by an estimating function 
approach~\citep{Jorgensen:2004,Bonat:2016}, where the quasi-score and Pearson estimating 
functions are adopted for estimation of regression and dispersion parameters, respectively. 
The estimating function approach allows us to extend the Poisson-Tweedie distributions to 
deal with underdispersed count data by allowing negative values for the dispersion 
parameter $\phi$. The Tweedie power parameter plays an important role in the 
Poisson-Tweedie family, since it is an index that distinguishes between important 
distributions, examples include Hermite ($p=0$), Neyman Type A ($p = 1$), 
P\'olya-Aeppli ($p=1.5$), negative binomial ($p=2$) and Poisson-inverse Gaussian ($p=3$). 
Furthermore, through the estimation of the Tweedie power parameter,
the Poisson-Tweedie family automatically adapts to highly 
skewed count data with excessive zeros, without the need to introduce zero-inflated or
hurdle components.

The Poisson-Tweedie family of distributions and its properties are introduced 
in Section~\ref{model}. In Section~\ref{estimation} we considered the estimating function
approach for parameter estimation and inference. Section \ref{simulation} presents
the main results of two simulation studies conducted to check the properties of the
estimating function derived estimators and explore the flexibility of the extended Poisson-Tweedie 
models to deal with underdispersed count data. The application of extended Poisson-Tweedie regression
models is illustrated in Section~\ref{results}.
Finally, discussions and directions for future work are given in Section \ref{discussion}.
The \texttt{R} implementation and the data sets are available in the 
supplementary material.

\section{Poisson-Tweedie: properties and regression models}~\label{model}

In this section, we derive the probability mass function and 
discuss the main properties of the Poisson-Tweedie distributions. 
Furthermore, we propose the extended Poisson-Tweedie regression model.
The Poisson-Tweedie distributions are Poisson Tweedie mixtures.
Thus, our initial point is an exponential dispersion model of the form
\begin{equation*}
\label{distri}
f_{Z}(z; \mu, \phi, p) = a(z,\phi,p) \exp\{(z\psi - k_p(\psi))/\phi\},
\end{equation*}
where $\mu = \mathrm{E}(Z) = k^{\prime}_p(\psi)$ is the mean, 
$\phi > 0$ is the dispersion parameter, $\psi$ is the canonical 
parameter and $k_p(\psi)$ is the cumulant function. 
The variance is given by $\mathrm{Var}(Z) = \phi V(\mu)$ 
where $V(\mu) = k^{\prime \prime}_p(\psi)$ is called the variance function. 
Tweedie densities are characterized by power variance functions of the 
form $V(\mu) = \mu^p$, where $p \in (-\infty  ,0] \cup [1,\infty)$ is 
the index determining the distribution.
For a Tweedie random variable $Z$, we write $Z \sim Tw_p(\mu, \phi)$.
The support of the distribution depends on the value of the
power parameter. For $p \geq 2$, $1 < p < 2$ and $p = 0$ the support 
corresponds to the positive, non-negative and real values, respectively.
In these cases $\mu \in \Omega$, where $\Omega$ is the convex support 
(i.e. the interior of the closed convex hull of the corresponding 
distribution support). Finally, for $p < 0$ the support again corresponds to the 
real values, however the expectation $\mu$ is positive. 

The function $a(z,\phi, p)$ cannot be written in a closed form, apart from 
the special cases corresponding to the Gaussian ($p = 0$), 
Poisson ($\phi = 1$ and $p = 1$), non-central gamma ($p = 3/2$), 
gamma ($p = 2$) and inverse Gaussian ($p = 3$) 
distributions~\citep{Jorgensen:1997,Bonat:2016a}. 
Another important case corresponds to the compound Poisson distributions,
obtained when $1 < p < 2$. The compound Poisson distribution is a frequent
choice for the modelling of non-negative data that has a probability mass at zero
and is highly right-skewed~\citep{Smyth:2002, Andersen:2016}.

The Poisson-Tweedie family is given by the following hierarchical specification
\begin{align}
\begin{split}
\label{conditional}
Y|Z &\sim \mathrm{Poisson}(Z) \\ 
Z &\sim \mathrm{Tw}_p(\mu, \phi). \nonumber
\end{split}
\end{align}
Here, we require $p \geq 1$, to ensure that $Z$ is non-negative.
In this case, the Poisson-Tweedie is an overdispersed factorial dispersion
model~\citep{Jorgensen:2016}. The probability mass function for $p > 1$ is given by
\begin{equation}
\label{pmf}
f(y;\mu,\phi,p) = \int_0^\infty \frac{z^y \exp{-z}}{y!} a(z,\phi,p) \exp\{(z\psi - k_p(\psi))/\phi\} dz.
\end{equation}
The integral (\ref{pmf}) has no closed-form apart of the special case 
corresponding to the negative binomial distribution, obtained when $p = 2$,
i.e. a Poisson gamma mixture. For $p=1$ the integral~(\ref{pmf}) 
is replaced by a sum and we have the Neyman Type A distribution. 
Further special cases include the 
Hermite $(p = 0)$, Poisson compound Poisson $(1 < p < 2)$,
factorial discrete positive stable $(p > 2)$ and Poisson-inverse Gaussian $(p = 3)$ 
distributions~\citep{Jorgensen:2016,Kokonendji:2004}.

\begin{figure}
\centering
\includegraphics{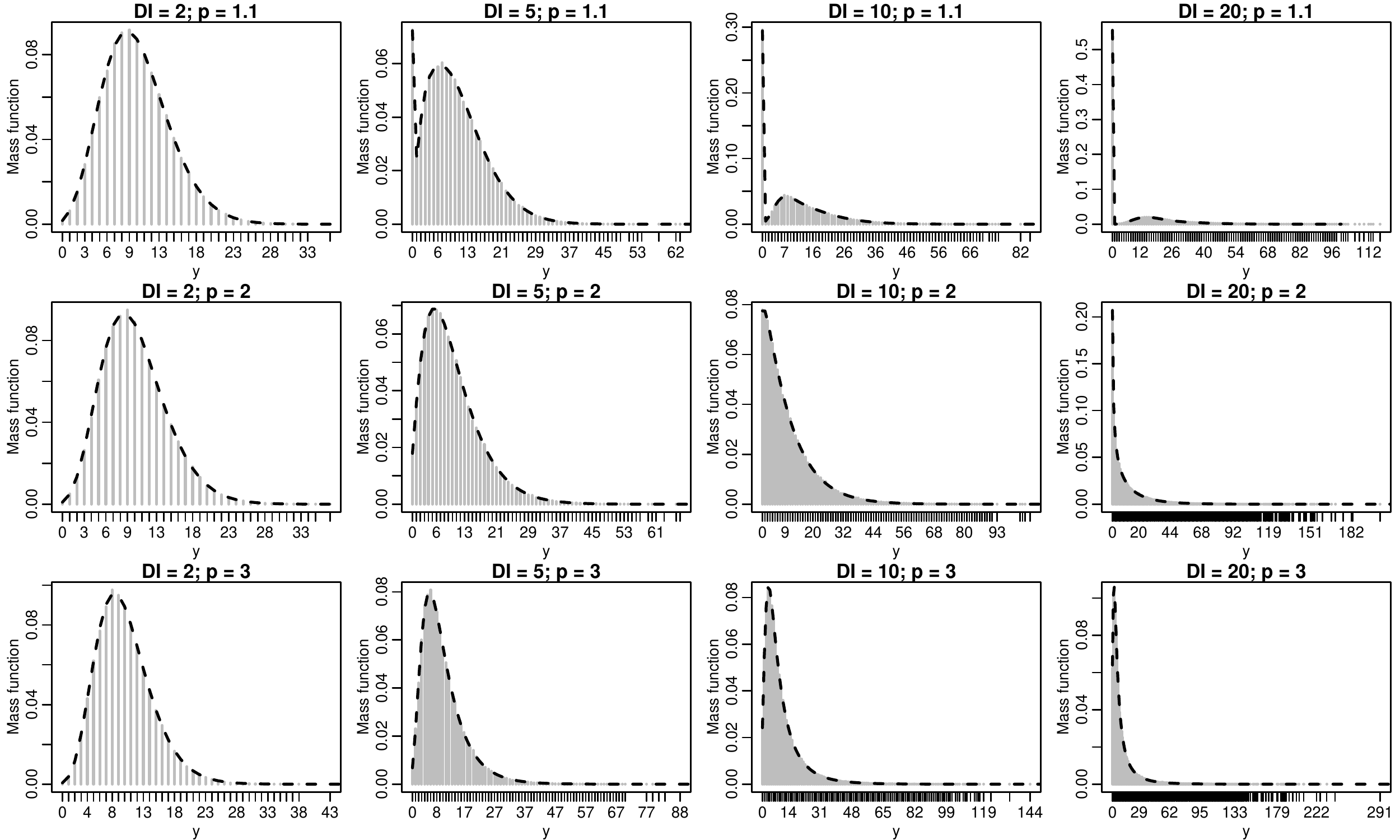}
\caption{Empirical (gray) and approximated (black) Poisson-Tweedie 
probability mass function by values of the dispersion index (DI) 
and Tweedie power parameter.}
\label{fig:shape}
\end{figure}

Simulation from Poisson-Tweedie distributions is easy because of the availability of
good simulation procedures for Tweedie distributions~\citep{tweedie:2013}. This also
makes it easy to approximate the integral (\ref{pmf}) using Monte Carlo integration,
since the Tweedie family is a natural proposal distribution. 
Alternatively, we can evaluate the integral using the Gauss-Laguerre method. 
Figure~\ref{fig:shape} presents the empirical probability mass function for some 
Poisson-Tweedie distributions computed based on a random sample of size $100,000$ (gray). 
Additionally, we display an approximation for the probability mass function (black line) 
obtained by Monte Carlo integration. We considered different values of the Tweedie power 
parameter ($p = 1.1$, $2$, $3$) combined with different values of the dispersion index 
($\mathrm{DI} =2,\,5,\,10,\,20$), which is defined by 
$$\mathrm{DI} = \mathrm{Var}(Y)/\mathrm{E}(Y).$$ 
In all scenarios the expectation $\mu$ was fixed at $10$.

\setkeys{Gin}{width=0.99\textwidth}

Figure~\ref{fig:shape} show that in the small dispersion case ($\mathrm{DI} = 2$) 
the shape of the probability mass functions is quite similar for the different values of 
the power parameter. However, when the dispersion index increases the 
differences become more marked. For $p = 1.1$ the overdispersion is clearly 
attributable to zero-inflation, while for $p = 3$ the overdispersion is due to the
heavy tail. The negative binomial case $(p = 2)$ is a critical point, 
where the distribution changes from zero-inflated to heavy-tailed. 
The results in Figure~\ref{fig:shape} also show that the Monte Carlo 
method provides a reasonable approximation for the probability mass 
function for all Poisson-Tweedie distributions. 

\begin{figure}
\centering
\includegraphics{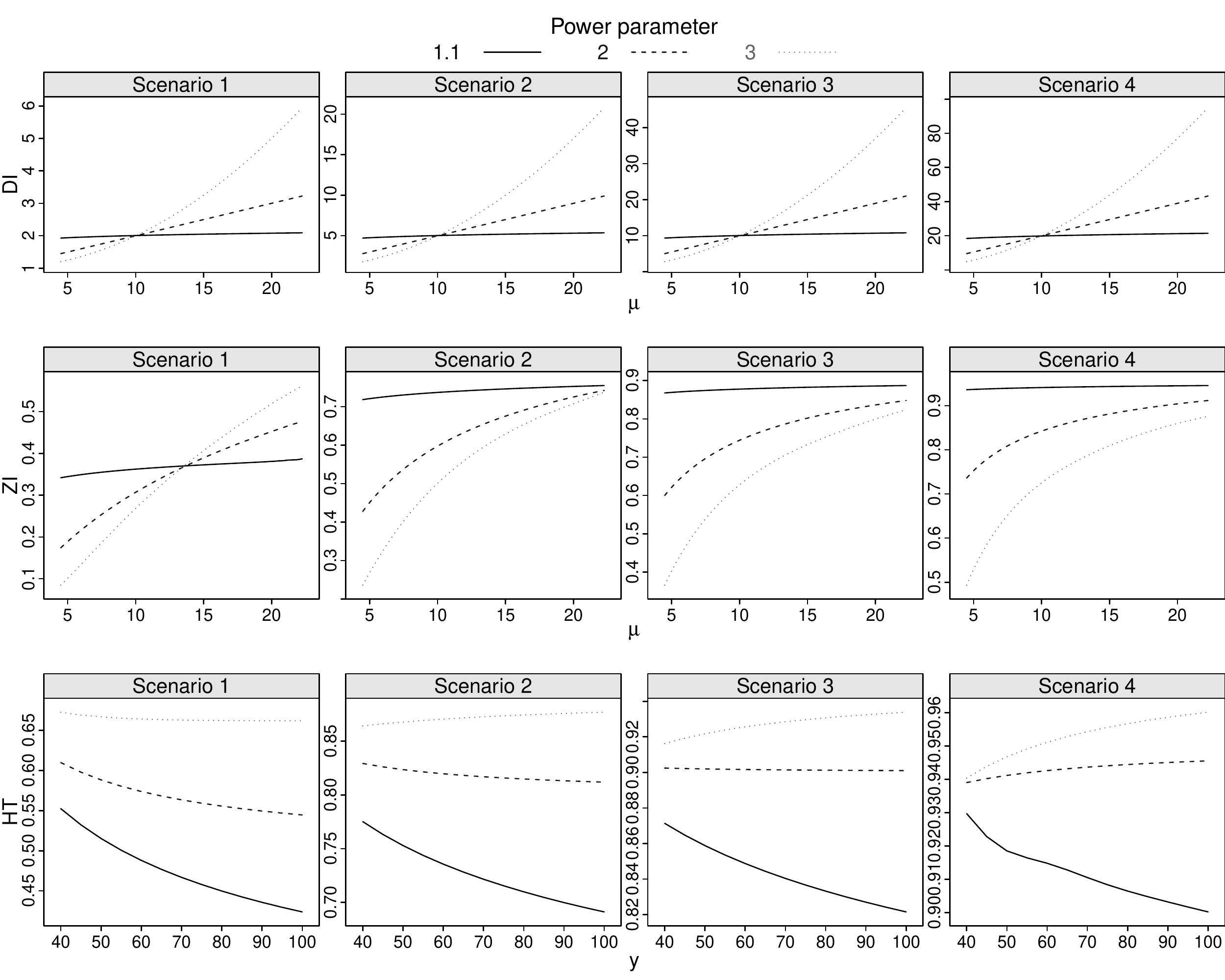}
\caption{Dispersion~$\mathrm{(DI)}$ and zero-inflation~$\mathrm{(ZI)}$ indices as a function of $\mu$ by 
simulation scenarios and Tweedie power parameter values. Heavy tail index (HT) for 
some extreme values of the random variable $Y$ by simulation scenarios 
and Tweedie power parameter values.}
\label{fig:index}
\end{figure}

In order to further explore the flexibility of the Poisson-Tweedie distributions, 
we introduce indices for zero-inflation
$$\mathrm{ZI} = 1 + \frac{\log \mathrm{P}(Y = 0)}{\mathrm{E}(Y)}$$ 
and a heavy tail
$$\mathrm{HT} = \frac{\mathrm{P}(Y=y+1)}{\mathrm{P}(Y=y)}\quad \text{for} \quad y \to \infty.$$ 
These indices are defined in relation to the Poisson distribution. 
The zero-inflated index is easily interpreted, since $\mathrm{ZI} < 0$ 
indicates zero-deflation, $\mathrm{ZI} = 0$ corresponds to no excess of zeroes, 
and $\mathrm{ZI} > 0$ indicates zero-inflation. 
Similarly, $\mathrm{HT} \to 1$ when $y \to \infty$ indicates a heavy tail 
distribution (for a Poisson distribution $\mathrm{HT} \to 0$ when $y \to \infty$). 
Figure~\ref{fig:index} presents the dispersion and 
zero-inflation indices as a function of the expected values $\mu$ for 
different values of the dispersion and Tweedie power parameters.
The expected values are defined by $\mu_i = \exp\{\log(10) + 0.8x_i\}$ 
where $x_i$ is a sequence of length $100$ from $-1$ to $1$.
We also present the heavy tail index for some extreme values of the 
random variable. The dispersion parameter was fixed in order to have 
$\mathrm{DI} = 2, 5, 10$ and $20$ when the mean equals $10$.
We refer to these different cases as simulation scenarios $1$ to $4$, respectively.

\setkeys{Gin}{width=0.99\textwidth}

The indices presented in Figure~\ref{fig:index} show that for small 
values of the power parameter the Poisson-Tweedie distribution is 
suitable to deal with zero-inflated count data. In that case, 
the $\mathrm{DI}$ and $\mathrm{ZI}$ are almost not dependent on the values 
of the mean. However, the $\mathrm{HT}$ decreases as the mean increases. 
On the other hand, for large values of the power parameter the 
$\mathrm{HT}$ increases with increasing mean, showing that the model 
is specially suitable to deal with heavy-tailed count data. 
In this case, the $\mathrm{DI}$ and $\mathrm{ZI}$ increase quickly as 
the mean increases giving an extremely overdispersed model for 
large values of the mean. In general, the $\mathrm{DI}$ and $\mathrm{ZI}$ 
are larger than one and zero, respectively, which, of course, shows
that the corresponding Poisson-Tweedie distributions cannot deal with underdispersed 
and zero-deflated count data.

In spite of the integral (\ref{pmf}) having no closed-form, the first two 
moments (mean and variance) of the Poisson-Tweedie family can 
easily be obtained. 
This fact motivates us to specify a model by using 
only second-order moment assumptions.
Consider a cross-sectional dataset, $(y_i, \boldsymbol{x}_i)$, $i = 1, \ldots, n$, where
$y_i$'s are i.i.d. realizations of $Y_i$ according to
$Y_i \sim \mathrm{PTw_p}(\mu_i, \phi)$ and $g(\mu_i) = \eta_i = \boldsymbol{x}_i^{\top} \boldsymbol{\beta}$,
where $\boldsymbol{x}_i$ and $\boldsymbol{\beta}$ are ($Q \times 1$) vectors of 
known covariates and unknown regression parameters, respectively. 
It is straightforward to show by using the factorial cumulant generating function~\citep{Jorgensen:2016} that 
\begin{align}
\begin{split}
\label{conditional}
\mathrm{E}(Y_i) = \mu_i = g^{-1}(\boldsymbol{x}_i^{\top} \boldsymbol{\beta}) \\
\mathrm{Var}(Y_i) = C_i = \mu_i + \phi \mu_i^p, 
\end{split}
\end{align}
where $g$ is a standard link function, for which here we adopt the logarithm link function. 
The Poisson-Tweedie regression model is parametrized by 
$\boldsymbol{\theta} = (\boldsymbol{\beta}^\top, \boldsymbol{\lambda}^\top = (\phi, p)^\top)^\top$.
Note that, based on second-order moment assumptions, the only restriction to have a proper model is that
$\mathrm{Var}(Y_i) > 0$, thus $$\phi > - \mu^{(1-p)}_i,$$ which shows that at least at some extent negative
values for the dispersion parameter are allowed. Thus, the Poisson-Tweedie model can be extended to deal
with underdispersed count data, however, in doing so the associated probability mass functions do not exist.

\setkeys{Gin}{width=0.99\textwidth}
\begin{figure}[h]
\centering
\includegraphics{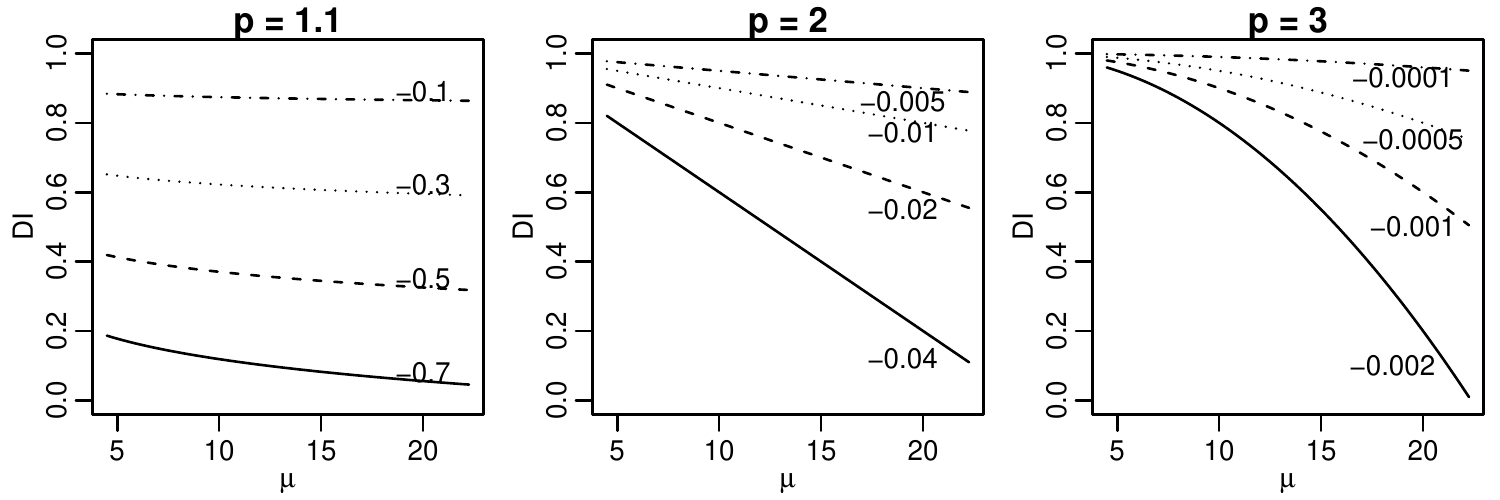}
\caption{Dispersion index as a function of $\mu$ by dispersion and Tweedie power parameter values.}
\label{fig:under}
\end{figure}

Figure~\ref{fig:under} presents the $\mathrm{DI}$ as a function of the mean
for different values of the Tweedie power parameter and negative values for the dispersion parameter.
As expected for negative values of the dispersion parameter the $\mathrm{DI}$ gives values smaller than $1$,
indicating underdispersion. We also note that, as the mean increases the $\mathrm{DI}$ decreases slowly 
for small values of the Tweedie power parameter and faster for larger values of the Tweedie power parameter. 
This shows that the range of negative values allowed for the dispersion parameter decreases rapidly as
the value of the Tweedie power parameter increases. Thus, for underdispersed data, we expect small values for
the Tweedie power parameter. Furthermore, the second-order moment assumptions also allow us to eliminate the non-trivial
restriction on the parameter space of the Tweedie power parameter. This makes it possible to estimate values between $0$
and $1$ where the corresponding Tweedie distribution does not exist. Table~\ref{tab:model} presents the main special cases
and the dominant features of the Poisson-Tweedie models according to the values of the dispersion and power parameters.

\begin{table}[h]
\centering
\caption{Reference models and dominant features by dispersion and power parameter values.}
\label{tab:model}
\begin{tabular}{llll} \hline
Reference Model          & Dominant features                    & Dispersion       & Power   \\ \hline
Poisson                  & Equi                                 & $\phi = 0$       & $-$      \\
Hermite                  & Over, Under                           & $\phi\lessgtr 0$ & $p = 0$   \\
Neyman Type A            & Over, Under, Zero-inflation          & $\phi\lessgtr 0$ & $p = 1$ \\
\textit{Poisson compound Poisson} & Over, Under,  Zero-inflation & $\phi\lessgtr 0$ & $1 < p < 2$ \\
P\'olya-Aeppli           & Over, Under, Zero-inflation          & $\phi\lessgtr 0$ & $p = 1.5$ \\
Negative binomial        & Over, Under                          & $\phi\lessgtr 0$ & $p = 2$ \\
\textit{Poisson positive stable}  & Over, heavy tail            & $\phi > 0$       & $p > 2$ \\
Poisson-inverse Gaussian & Over, heavy tail                     & $\phi > 0$       & $p = 3$ \\ \hline
\end{tabular}
\end{table}

\section{Estimation and Inference}~\label{estimation}

We shall now introduce the estimating function approach using terminology and results
from \citet{Jorgensen:2004} and \citet{Bonat:2016}. 
The estimating function approach adopted in this paper combines the quasi-score and 
Pearson estimating functions for estimation of regression and dispersion parameters, 
respectively. The quasi-score function for $\boldsymbol{\beta}$ has the following form,

\begin{equation*}
\psi_{\boldsymbol{\beta}}(\boldsymbol{\beta}, \boldsymbol{\lambda}) = \left (\sum_{i=1}^n \frac{\partial \mu_i}{\partial \beta_1}C^{-1}_i(y_i - \mu_i)^\top, \ldots, \sum_{i=1}^n \frac{\partial \mu_i}{\partial \beta_Q}C^{-1}_i(y_i - \mu_i)^\top  \right )^\top,
\end{equation*}
where $\partial \mu_i/\partial \beta_j = \mu_i x_{ij}$ for $j = 1, \ldots, Q$.
The entry $(j,k)$ of the $Q \times Q$ sensitivity matrix for $\psi_{\boldsymbol{\beta}}$ is given by

\begin{equation}
\label{Sbeta}
\mathrm{S}_{\beta_{jk}} = \mathrm{E}\left ( \frac{\partial}{\partial \beta_k} \psi_{\beta_j}(\boldsymbol{\beta}, \boldsymbol{\lambda})  \right ) = -\sum_{i=1}^n \mu_i x_{ij} C^{-1}_i x_{ik} \mu_i.
\end{equation}
In a similar way, the entry $(j,k)$ of the $Q \times Q$ variability matrix for $\psi_{\boldsymbol{\beta}}$ is given by
\begin{equation*}
\label{Vbeta}
\mathrm{V}_{\beta_{jk}} = \mathrm{Var}(\psi_{\boldsymbol{\beta}}(\boldsymbol{\beta}, \boldsymbol{\lambda})) = \sum_{i=1}^n \mu_i x_{ij} C^{-1}_i x_{ik} \mu_i.
\end{equation*}

Following \citet{Jorgensen:2004, Bonat:2016}, the Pearson estimating function for the dispersion parameters has the following form,

\begin{equation*}
\label{Pearson}
\psi_{\boldsymbol{\lambda}}(\boldsymbol{\lambda}, \boldsymbol{\beta}) = \left (\sum_{i=1}^n \boldsymbol{W}_{i\phi} \left [ (y_i - \mu_i)^2 - C_i \right ]^\top, \sum_{i=1}^n \boldsymbol{W}_{i p} \left [ (y_i - \mu_i)^2 - C_i \right ]^\top  \right )^\top,
\end{equation*}
where $\boldsymbol{W}_{i\phi} = - \partial C^{-1}_i/\partial \phi$ and $\boldsymbol{W}_{ip} = - \partial C^{-1}_i/\partial p$.
The Pearson estimating functions are unbiased estimating functions for $\boldsymbol{\lambda}$ 
based on the squared residuals $(y_i - \mu_i)^2$ with expected value $C_i$. 

The entry $(j,k)$ of the $2 \times 2$ sensitivity matrix for the dispersion parameters is given by
\begin{equation}
\label{Slambda}
\mathrm{S}_{\boldsymbol{\lambda}_{jk}} = \mathrm{E}\left ( \frac{\partial}{\partial \lambda_k}\psi_{\lambda_j}(\boldsymbol{\lambda}, \boldsymbol{\beta})  \right ) = -\sum_{i=1}^n \boldsymbol{W}_{i \lambda_j} C_i\boldsymbol{W}_{i\lambda_k}C_i, 
\end{equation}
where $\lambda_1$ and $\lambda_2$ denote either $\phi$ or $p$.

Similarly, the cross entries of the sensitivity matrix are given by
\begin{equation}
\label{Sbetalambda}
\mathrm{S}_{\beta_j \lambda_k} = \mathrm{E}\left ( \frac{\partial}{\partial \lambda_k}\psi_{\beta_j}(\boldsymbol{\beta}, \boldsymbol{\lambda})  \right ) = 0
\end{equation}
and
\begin{equation}
\label{Slambdabeta}
\mathrm{S}_{\lambda_j \beta_k} = \mathrm{E}\left ( \frac{\partial}{\partial \beta_k}\psi_{\lambda_j}(\boldsymbol{\lambda}, \boldsymbol{\beta})  \right ) = -\sum_{i=1}^n \boldsymbol{W}_{i\lambda_j}C_i\boldsymbol{W}_{i\beta_k}C_i,
\end{equation}
where $\boldsymbol{W}_{i\beta_k} = -\partial C_i^{-1}/\partial \beta_k$. 
Finally, the joint sensitivity matrix for the parameter vector $\boldsymbol{\theta}$ is given by
\begin{equation*}
\mathrm{S}_{\boldsymbol{\theta}} = \begin{pmatrix}
\mathrm{S}_{\boldsymbol{\beta}} & \boldsymbol{0} \\ 
\mathrm{S}_{\boldsymbol{\lambda}\boldsymbol{\beta}} & \mathrm{S}_{\boldsymbol{\lambda}}
\end{pmatrix},
\end{equation*}
whose entries are defined by equations (\ref{Sbeta}), (\ref{Slambda}), (\ref{Sbetalambda}) and (\ref{Slambdabeta}).

We now calculate the asymptotic variance of the estimating function estimators denoted by $\boldsymbol{\hat{\theta}}$, as
obtained from the inverse Godambe information matrix, whose general form for a vector of parameter $\boldsymbol{\theta}$ is 
$J^{-1}_{\boldsymbol{\theta}} = \mathrm{S}^{-1}_{\boldsymbol{\theta}} \mathrm{V}_{\boldsymbol{\theta}} \mathrm{S}^{-\top}_{\boldsymbol{\theta}}$, where $-\top$ denotes inverse transpose. 
The variability matrix for $\boldsymbol{\theta}$  has the form
\begin{equation}
\label{VTHETA}
\mathrm{V}_{\boldsymbol{\theta}} = \begin{pmatrix}
\mathrm{V}_{\boldsymbol{\beta}} & \mathrm{V}_{\boldsymbol{\beta}\boldsymbol{\lambda}} \\ 
\mathrm{V}_{\boldsymbol{\lambda}\boldsymbol{\beta}} & \mathrm{V}_{\boldsymbol{\lambda}}
\end{pmatrix},
\end{equation}
where $\mathrm{V}_{\boldsymbol{\lambda}\boldsymbol{\beta}} = \mathrm{V}^{\top}_{\boldsymbol{\beta}\boldsymbol{\lambda}}$ and $\mathrm{V}_{\boldsymbol{\lambda}}$ depend on the third and fourth moments of $Y_i$, respectively. In order to avoid this dependence on higher-order moments, we propose to use the empirical versions of $\mathrm{V}_{\boldsymbol{\lambda}}$ and $\mathrm{V}_{\boldsymbol{\lambda}\boldsymbol{\beta}}$ as given by

\begin{equation*}
\tilde{\mathrm{V}}_{\lambda_{jk}} = \sum_{i=1}^n \psi_{\lambda_j}(\boldsymbol{\lambda}, \boldsymbol{\beta})_i\psi_{\lambda_k}(\boldsymbol{\lambda}, \boldsymbol{\beta})_i \quad \text{and} \quad \tilde{\mathrm{V}}_{\lambda_j \beta_k} = \sum_{i=1}^n \psi_{\lambda_j}(\boldsymbol{\lambda}, \boldsymbol{\beta})_i\psi_{\beta_k}(\boldsymbol{\lambda}, \boldsymbol{\beta})_i.
\end{equation*}
Finally, the asymptotic distribution of $\boldsymbol{\hat{\theta}}$ is given by
\begin{equation*}
\boldsymbol{\hat{\theta}} \sim \mathrm{N}(\boldsymbol{\theta}, \mathrm{J}_{\boldsymbol{\theta}}^{-1}), \quad \text{where} \quad
J^{-1}_{\boldsymbol{\theta}} = \mathrm{S}^{-1}_{\boldsymbol{\theta}} \mathrm{V}_{\boldsymbol{\theta}} \mathrm{S}^{-\top}_{\boldsymbol{\theta}}.
\end{equation*} 

To solve the system of equations $\psi_{\boldsymbol{\beta}} = \boldsymbol{0}$ and $\psi_{\boldsymbol{\lambda}} = \boldsymbol{0}$ \citet{Jorgensen:2004} proposed the modified chaser algorithm, defined by
\begin{eqnarray*}
\label{chaser}
\boldsymbol{\beta}^{(i+1)} &=& \boldsymbol{\beta}^{(i)} - \mathrm{S}_{\boldsymbol{\beta}}^{-1} \psi_{\boldsymbol{\beta}}(\boldsymbol{\beta}^{(i)}, \boldsymbol{\lambda}^{(i)}) \nonumber \\
\boldsymbol{\lambda}^{(i+1)} &=& \boldsymbol{\lambda}^{(i)} - \alpha \mathrm{S}_{\boldsymbol{\lambda}}^{-1} \psi_{\boldsymbol{\lambda}}(\boldsymbol{\beta}^{(i+1)}, \boldsymbol{\lambda}^{(i)}).
\end{eqnarray*}
The modified chaser algorithm uses the insensitivity property (\ref{Sbetalambda}), which allows us to use two
separate equations to update $\boldsymbol{\beta}$ and $\boldsymbol{\lambda}$. We introduce the tuning constant, $\alpha$, to control the step-length. A similar version of this algorithm was used by \citet{Bonat:2016a} for estimation and inference in the context of Tweedie regression
models. Furthermore, this algorithm is a special case of the flexible algorithm presented by \citet{Bonat:2016} in the context of multivariate covariance generalized linear models. Hence, estimation for the Poisson-Tweedie model is easily implemented in \texttt{R} through the \texttt{mcglm}~\citep{Bonat:2015a} package.

\section{Simulation studies}~\label{simulation}
In this section we present two simulation studies designed to
explore the flexibility of the extended Poisson-Tweedie models to deal with
over and underdispersed count data.

\subsection{Fitting extended Poisson-Tweedie models to overdispersed data}

In this first simulation study we designed $12$ simulation scenarios 
to explore the flexibility of the extended Poisson-Tweedie model
to deal with overdispersed count data. For each setting,
we considered four different sample sizes,
$100$, $250$, $500$ and $1000$, generating $1000$ datasets in each case. 
We considered three values 
of the Tweedie power parameter, $1.1$, $2$ and $3$, combined with four different degrees of 
dispersion as measured by the dispersion index. In the case of $p=1.1$,
the dispersion parameter was fixed at $\phi = 0.8, 3.2, 7.2$ and $15$.
Similarly, for $p = 2$ and $p = 3$ the dispersion parameter was fixed
at $\phi = 0.1, 0.4, 0.9,1.9$ and $\phi = 0.01, 0.04, 0.09, 1.9$, 
respectively. These values were chosen so that when the mean is $10$ the 
dispersion index takes values of $2$, $5$, $10$ and $20$, respectively.
The probability mass function of the Poisson-Tweedie distribution
for each parameter combination is as presented in Figure~\ref{fig:shape}.

\begin{figure}
\centering
\includegraphics{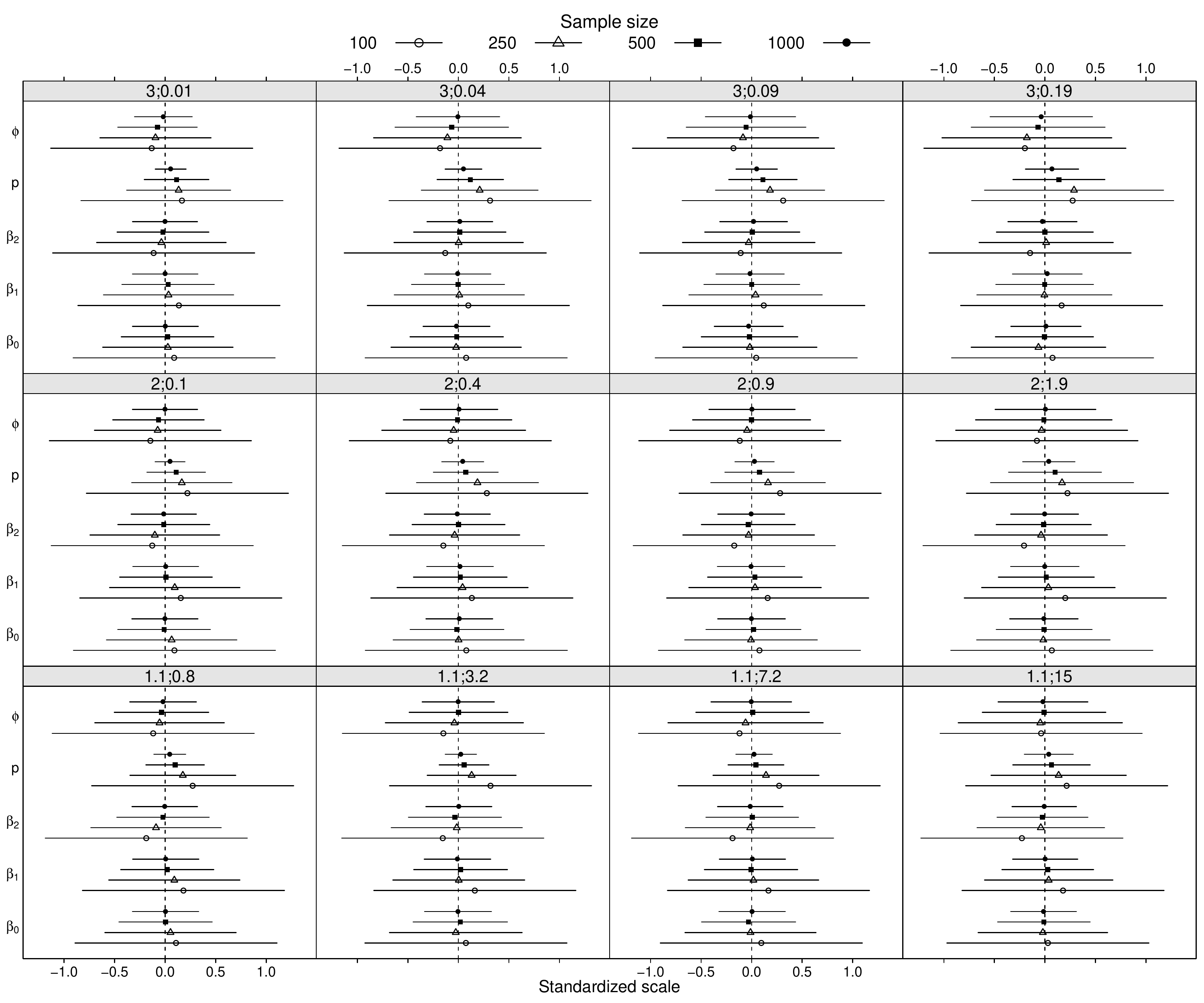}
\caption{Average bias and  confidence intervals on a standardized scale by sample size 
and simulation scenario.}
\label{fig:simul}
\end{figure}

In order to have a regression model structure, we specified the mean vector as
$\mu_i = \exp\{\log(10) + 0.8 x_{1i} - 1 x_{2i}\}$, where $x_{1i}$ is
a sequence from $-1$ to $1$ with length equals to the sample size.
Similarly, the covariate $x_{2i}$ is a categorical covariate with two
levels $(0$ and $1)$ and length equals sample size.
Figure~\ref{fig:simul}
shows the average bias plus and minus the average standard error for the
parameters under each scenario. The scales are standardized for
each parameter by dividing the average bias and the limits of the  confidence
intervals by the standard error obtained for the sample of size $100$.

\setkeys{Gin}{width=0.99\textwidth}

\begin{figure}
\centering
\includegraphics{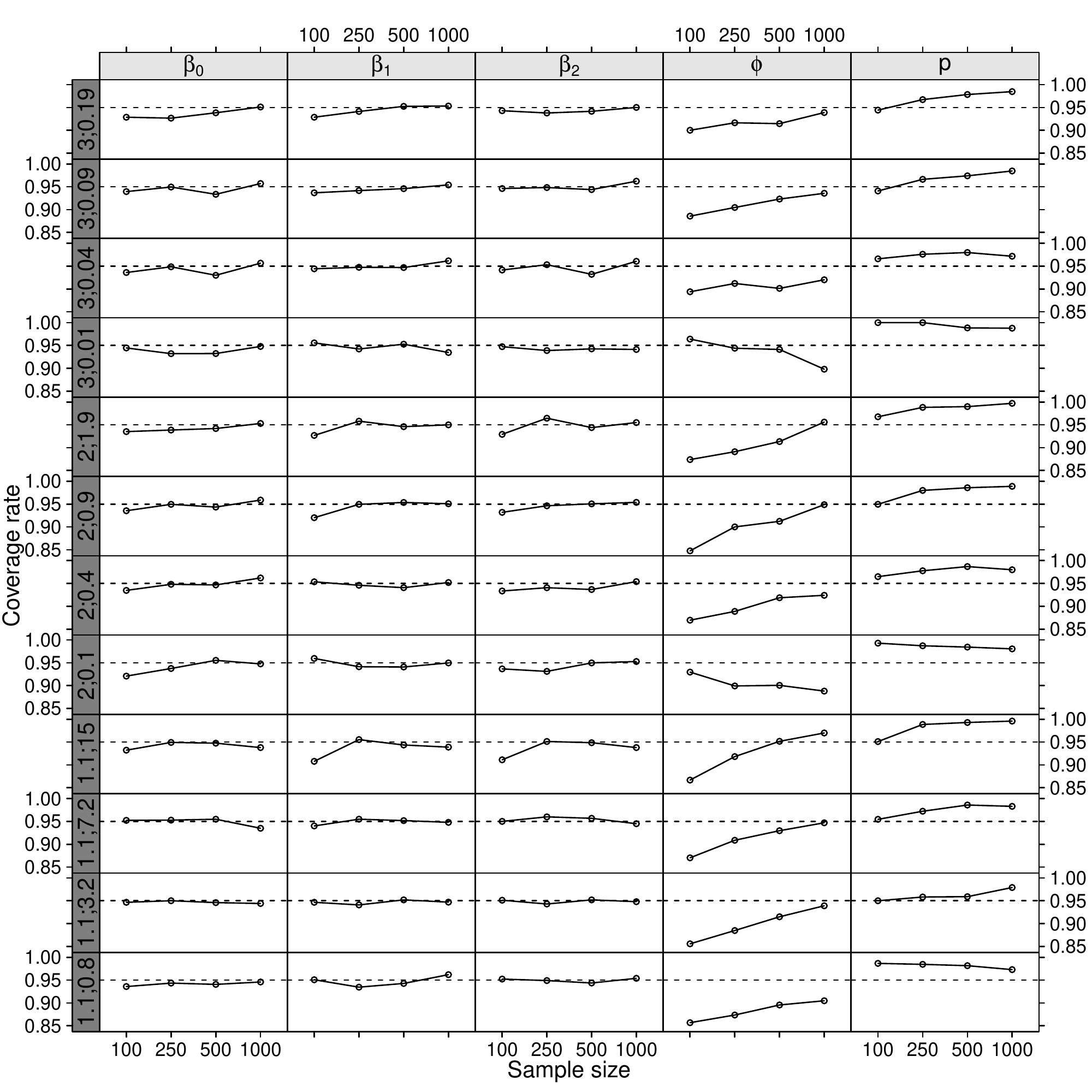}
\caption{Coverage rate for each parameter by sample size and simulation scenarios.}
\label{fig:coverage}
\end{figure}

The results in Figure~\ref{fig:simul} show that for all simulation 
scenarios both the average bias and standard errors tend to $0$ as the 
sample size is increased. This shows the consistency and unbiasedness of 
the estimating function estimators. Figure~\ref{fig:coverage} presents 
the confidence interval coverage rate by sample size and simulation scenarios.

\setkeys{Gin}{width=0.99\textwidth}

The results presented in Figure~\ref{fig:coverage} show that for the 
regression parameters the empirical coverage rates are close to the nominal 
level of $95\%$  for all sample sizes and simulation scenarios. 
For the dispersion parameter and a small sample size the empirical coverage 
rates are slightly lower than the nominal level, however, they become
closer for large samples. On the other hand, for the power parameter the 
empirical coverage rates were slightly larger than the nominal level, 
for all sample sizes and simulation scenarios.

\subsection{Fitting extended Poisson-Tweedie models to underdispersed data}

As discussed in Section~\ref{model}, the extended Poisson-Tweedie model 
can deal with underdispersed count data by allowing negative values for 
the dispersion parameter. However, in that case there is no probability 
mass function associated with the model. 
Consequently, it is impossible to use such a model to simulate 
underdispersed data. Thus, we simulated data sets from the 
COM-Poisson~\citep{Sellers:2010} and Gamma-Count~\citep{Zeviani:2014} 
distributions. Such models are well known in the literature for their 
ability to model underdispersed data.

Following the parametrization used by~\citet{Sellers:2010},  
$Y \sim CP(\lambda, \nu)$ denotes a COM-Poisson distributed random variable.
Similarly, we write $Y \sim GC(\lambda, \nu)$ for a Gamma-Count distributed
random variable. For both distributions the additional 
parameter $\nu$ controls the dispersion structure, with values
larger than $1$ indicating underdispersed count data.
An inconvenience of the COM-Poisson and Gamma-Count 
regression models as proposed by~\citet{Sellers:2010} 
and~\citet{Zeviani:2014}, respectively,  is that the regression structure is not linked 
to a function of $E(Y)$ as is usual in the generalized linear models framework.
To overcome this limitation and obtain  parameters that are interpretable in the 
usual way, i.e. related directly to a function of $E(Y)$, we take an alternative 
approach based on simulation. The procedure consisted of specifying the 
$\lambda$ parameter using a regression structure, 
$\lambda_i =\exp\{\lambda_0 + \lambda_1 x_1\}$ for $i = 1, \ldots, n$ 
where $n$ denotes the sample size and $x_1$ is a sequence from $-1$ to 
$1$ and length $n$.
For each value of $\lambda$ we simulate $1000$ values and compute the
empirical mean and variance. We denote these quantities by 
$\widehat{E(Y)}$ and $\widehat{var(Y)}$. Then, we fitted two non-linear models 
specified as
$\widehat{E(Y)} = \exp(\beta_0 + \beta_1 x_1)$ and 
$\widehat{var(Y)} = \widehat{E(Y)} + \phi \widehat{E(Y)}^p$.
From these fits, we obtained the expected values of the regression, 
dispersion and Tweedie power parameters. 

We designed four simulation scenarios by introducing different 
degrees of underdispersion in the data sets. 
The parameter $\nu$ was fixed at the values 
$\nu = 2, 4, 6$ and $8$ for both distributions. 
In the COM-Poisson case we took $\lambda_0 = 8$ and $\lambda_1 = 4$ and for the 
Gamma-Count case we fixed $\lambda_0 = 2$ and $\lambda_1 = 1$.
It is important to highlight that for all of these selected values the expected 
value of the dispersion parameter $\phi$ is negative. 
The particular values depend on $\lambda_0$, $\lambda_1$ and $\nu$ 
and are presented for both distributions in Table~\ref{S1}.

\begin{table}[h]
\centering
\caption{Corresponding values of $\beta_0$, $\beta_1$, $\phi$ and $p$ depending on the values of 
$\lambda_0$, $\lambda_1$ and $\nu$ for the COM-Poisson and Gamma-Count distributions.}
\label{S1}
\begin{tabular}{cccccccc} \hline
\multicolumn{7}{c}{COM-Poisson}                                       \\ \hline
$\nu$ & $\lambda_0$ & $\lambda_1$ & $\beta_0$ & $\beta_1$ & $\phi$   & $p$     \\ \hline
$2$   & $8$         & $4$         & $3.995$   & $2.004$   & $-0.485$ & $1.008$ \\
$4$   & $8$         & $4$         & $1.941$   & $1.047$   & $-0.714$ & $1.014$ \\
$6$   & $8$         & $4$         & $1.206$   & $0.744$   & $-0.790$ & $1.020$ \\
$8$   & $8$         & $4$         & $0.803$   & $0.602$   & $-0.821$ & $1.036$ \\ \hline
\multicolumn{7}{c}{Gamma-Count}                                      \\ \hline
$2$   & $2$         & $1$         & $1.962$   & $1.028$   & $-0.429$ & $1.045$ \\ 
$4$   & $2$         & $1$         & $1.943$   & $1.042$   & $-0.682$ & $1.003$ \\
$6$   & $2$         & $1$         & $1.936$   & $1.048$   & $-0.779$ & $1.019$ \\
$8$   & $2$         & $1$         & $1.932$   & $1.051$   & $-0.820$ & $1.020$ \\ \hline
\end{tabular}
\end{table}

\begin{figure}
\centering
\includegraphics{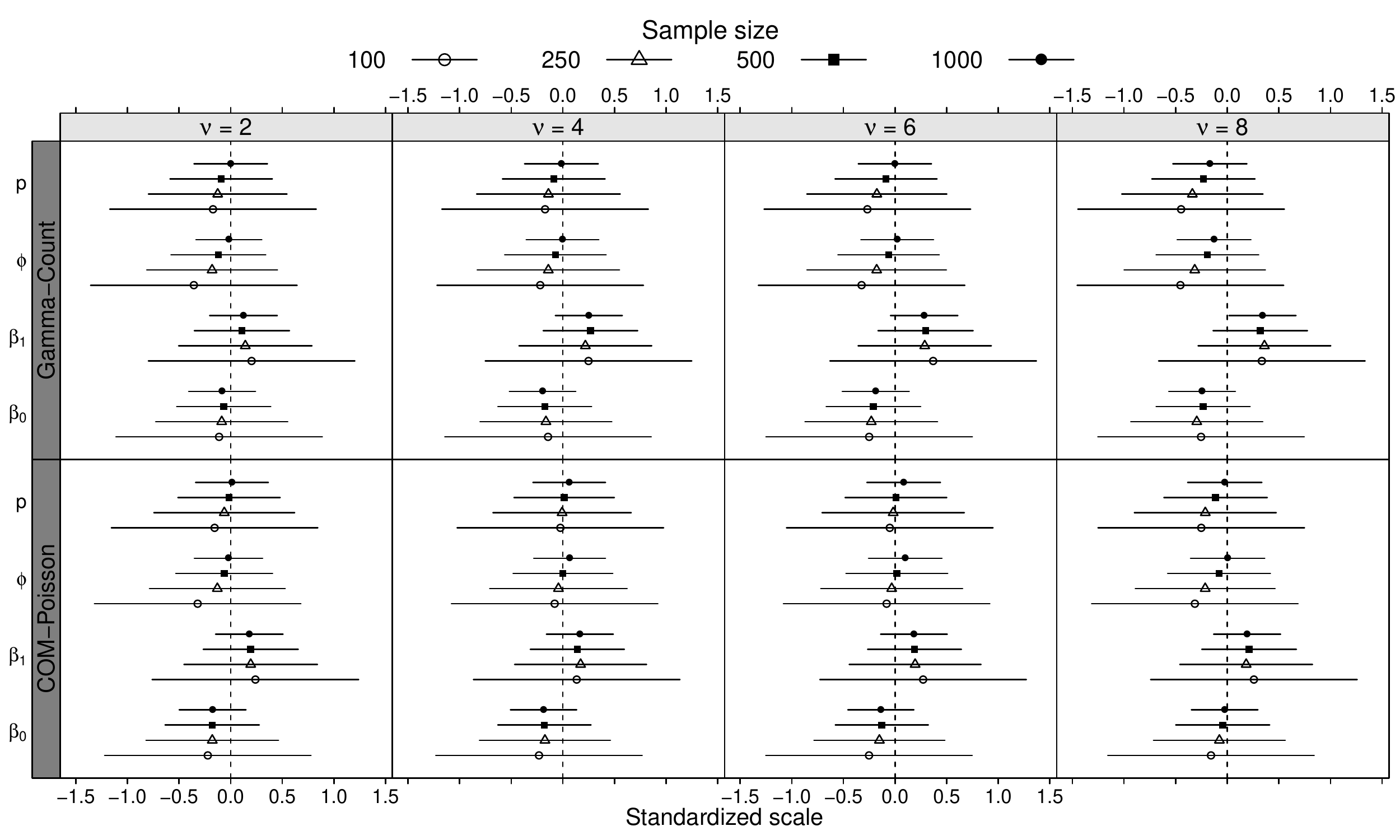}
\caption{Average bias and confidence interval on a standardized scale 
by sample size and simulation scenario.}
\label{fig:simulUnder}
\end{figure}

For each setting, we generated $1000$ data sets for four different sample sizes 
$100$, $250$, $500$ and $1000$. The extended Poisson-Tweedie model was
fitted using the estimating function approach presented in the 
Section~\ref{estimation}.
Figure~\ref{fig:simulUnder} shows the average bias plus and minus the 
average standard error for the parameters in each scenario. 
For each parameter the scales are standardized by  dividing the average 
bias and  limits of the confidence intervals by the standard error 
obtained for the sample of size $100$.

\setkeys{Gin}{width=0.99\textwidth}

The results in Figure~\ref{fig:simul} show that for all simulation 
scenarios, both the average bias and standard errors tend to $0$ as the 
sample size is increased for both dispersion and Tweedie power parameters. 
It shows the consistency of the estimating function estimators.
Concerning the regression parameters, in general the intercept~($\beta_0$) 
is underestimated, while the slope~($\beta_1$) is overestimated. 
The bias is larger for the Gamma-Count data with strong 
underdispersion~($\nu = 8$) case. However, it is still small in its 
magnitude.

\section{Data analyses}~\label{data}

In this section we present four examples to illustrate the 
application of the extended Poisson-Tweedie models. 
The data and the R scripts used for their analysis can be obtained \\
\texttt{http://www.leg.ufpr.br/doku.php/publications:papercompanions:ptw}.

\subsection{Data set 1: respiratory disease morbidity among children 
in Curitiba, Paran\'a, Brazil}
\begin{figure}[h]
\centering
\includegraphics{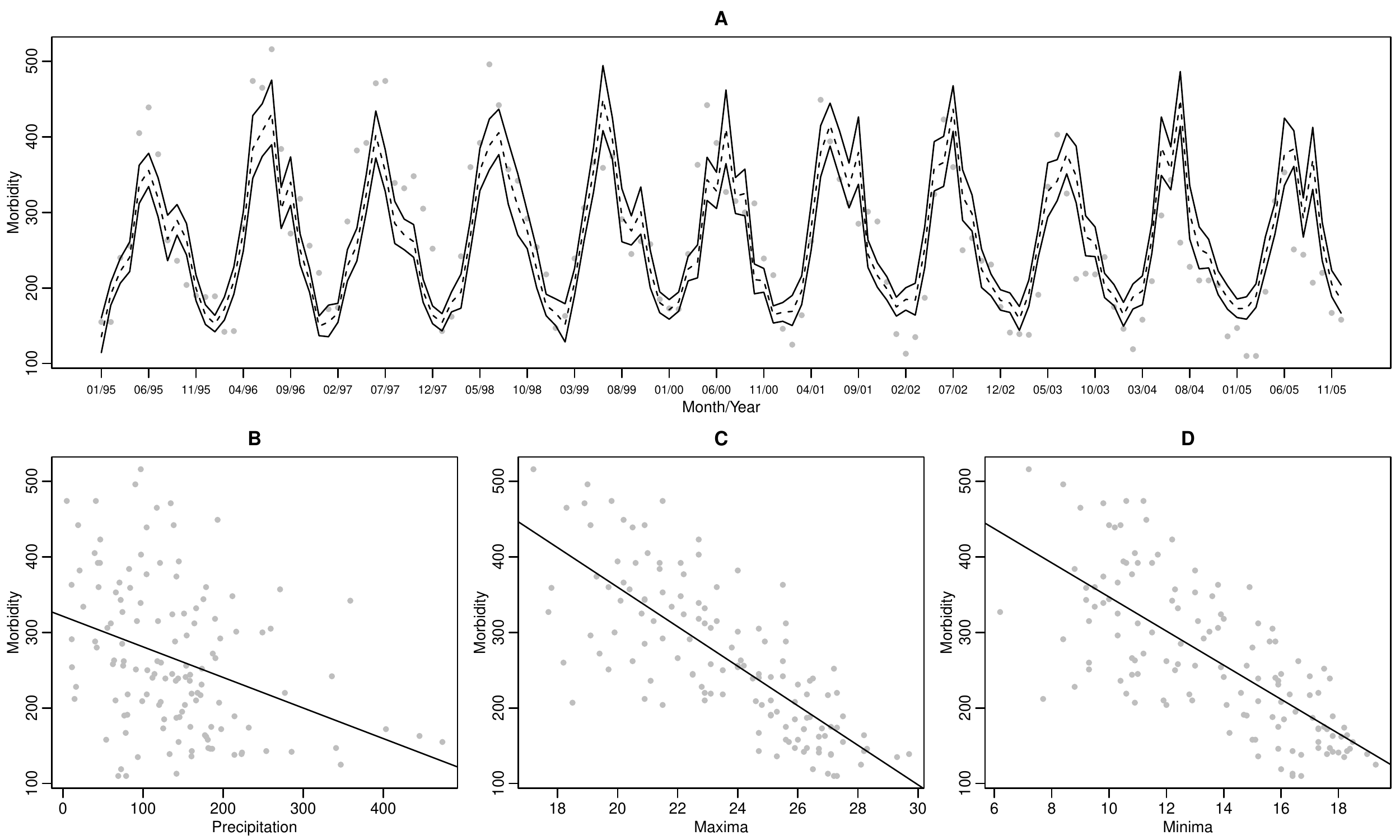}
\caption{Time series plot with fitted values~(A) and dispersion diagrams 
of the monthly morbidity by respiratory diseases against the covariates 
precipitation~(B), maximum temperature~(C) and minimum temperature~(D), 
with a simple linear fit indicated by the straight black lines.}
\label{fig:example1}
\end{figure}

The first example concerns monthly morbidity from respiratory diseases among 
$0$ to $4$ year old children in Curitiba, Paran\'a State, Brazil. 
The data were collected for the period from January $1995$ to December 
$2005$, corresponding to $132$ months. The main goal of the investigation 
was to assess the effect of three environmental covariates (precipitation, 
maximum and minimum temperatures) on the morbidity from respiratory diseases. 
Figure~\ref{fig:example1} presents a time series plot with fitted 
values~(A) and dispersion diagrams of the monthly morbidity from 
respiratory diseases against the covariates precipitation~(B), 
maximum temperature~(C) and minimum temperature~(D), 
with a simple linear fit indicated by the straight black lines.
\setkeys{Gin}{width=0.99\textwidth}
These plots indicate a clear seasonal pattern and the essentially
linear effect of all covariates (as suggested by the simple linear fits superimposed 
in Figure~\ref{fig:example1}). The linear predictor is expressed 
in terms of Fourier harmonics (seasonal variation) and the effect of the 
three environmental covariates. The logarithm of the population size was 
used as an offset. To compare the extended Poisson-Tweedie model with 
the usual Poisson log-linear model, Table~\ref{tab:example1} shows the corresponding 
estimates and standard errors~(SE), along with the ratios between the both model estimates and standard errors.

\begin{table}[h]
\centering
\caption{Data set $1$: Parameter estimates and standard errors (SE) for Poisson-Tweedie 
and Poisson models (first and second columns). Ratios between 
Poisson-Tweedie and Poisson estimates and standard errors (third column).}
\label{tab:example1}
\begin{tabular}{lrrr} \hline
\multirow{2}{*}{Parameter}     &\multicolumn{3}{c}{Estimates (SE)}  \\ \cline{2-4}
                            & Poisson-Tweedie       & Poisson               & Ratio  \\ \hline
\texttt{Intercept}          & $2.277~(0.304)^*$     & $2.226~(0.084)^*$  & $1.023~(3.598)$  \\
\texttt{cos(2*pi*Month/12)} & $-0.223~(0.056)^*$    & $-0.226~(0.016)^*$ & $0.985~(3.507)$   \\
\texttt{sin(2*pi*Month/12)} & $-0.093~(0.048)^*$    & $-0.073~(0.013)^*$ & $1.279~(3.562)$   \\
\texttt{Maxima}             & $-0.083~(0.017)^*$    & $-0.083~(0.005)^*$ & $1.057~(3.590)$   \\
\texttt{Minima}             & $0.039~(0.022)~$      & $0.034~(0.006)^*$  & $1.128~(3.592)$   \\
\texttt{Precipitation}      & $-0.001~(0.000)~$     & $-0.001~(0.000)^*$ & $0.978~(3.337)$   \\ \hline
$p$                         & $1.652~(0.423)~$      & $-$                   & $-$   \\
$\phi$                      & $0.293~(0.036)~$      & $-$                    & $-$    \\ \hline     
\end{tabular}
\end{table}

The results presented in Table~\ref{tab:example1} show that the estimates 
from the extended Poisson-Tweedie and Poisson models are similar. 
However, the standard errors from the extended Poisson-Tweedie model are
in general $3.5$ times larger than the ones from the Poisson model. 
This difference is explained by the dispersion structure. 
The dispersion parameter $\phi > 0$ indicates overdispersion, 
which implies that the standard errors obtained by the Poisson model 
are underestimated. The Poisson model gives evidence of a significant 
effect for all covariates, while the Poisson-Tweedie model only gives 
significant effects for the seasonal variation and the temperature 
maxima covariates.
The fitted values and $95\%$ confidence interval are shown in 
Figure~\ref{fig:example1}(A). The model captures the swing in the data 
and highlights the seasonal behaviour with high and low morbidity numbers 
around winter and summer months, respectively. The negative effect of 
the covariate temperature maxima agrees with the seasonal effects and 
the exploratory analysis presented in Figure~\ref{fig:example1}(C). 
The power parameter estimate with its corresponding standard error indicate 
that all Poisson-Tweedie models with $p \in [1,2]$ are suitable for this data set.
In particular, Neyman Type A, P\'olya-Aeppli and negative binomial distributions
can be good choices. 

\subsection{Data set 2: cotton bolls greenhouse experiment}
The second example relates to cotton boll production and is from  a completely randomized experiment
conducted in a greenhouse. The aim was to assess the effect of five artificial 
defoliation levels ($0\%$, $25\%$, $50\%$, $75\%$ and $100\%$) 
and five growth stages (vegetative, flower-bud, blossom, fig and cotton boll) 
on the number of cotton bolls.
There were five replicates of each treatment combination, giving a data set with $125$ observations. 
This data set was analysed in 
\citet{Zeviani:2014} using the Gamma-Count distribution, since there was clear evidence of underdispersion. Following~\citet{Zeviani:2014},
the linear predictor was specified by
$$ g(\mu_{ij}) = \beta_0 + \beta_{1j} \texttt{def}_i + \beta_{2j} \texttt{def}^2_i,$$ 
where $\mu_{ij}$ is the expected number of cotton bolls for the 
defoliation~(\texttt{def}) level $i = 1, \ldots, 5$ and 
growth stage $j = 1, \ldots, 5$, that is, we have a second order effect of defoliation 
in each growth stage. Table~\ref{tab:example2} presents the estimates and standard errors for 
the Poisson-Tweedie and standard Poisson models along, with the ratios
between the respective estimates and standard errors.

\begin{table}[h]
\centering
\caption{Data set $2$: Parameter estimates and standard errors (SE) for Poisson-Tweedie 
and Poisson models (first and second columns). Ratios between 
Poisson-Tweedie and Poisson estimates and standard errors (third column).}
\label{tab:example2}
\begin{tabular}{lrrrr} \hline
\multirow{2}{*}{Parameter}     &\multicolumn{3}{c}{Estimates (SE)} \\ \cline{2-4}
                            & Poisson-Tweedie    & Poisson             & Ratio  \\ \hline
\texttt{Intercept}          & $2.189~(0.030)^*$  & $2.190~(0.063)^*$   & $1.000~(0.471)$  \\
\texttt{vegetative:des}     & $0.438~(0.243)~$   & $0.437~(0.516)~$    & $1.003~(0.471)$   \\
\texttt{vegetative:des$^2$} & $-0.806~(0.274)^*$ & $-0.805~(0.584)~$   & $1.001~(0.469)$   \\
\texttt{flower bud:des}     & $0.292~(0.239)~$   & $0.290~(0.508)~$    & $1.007~(0.471)$   \\
\texttt{flower bud:des$^2$} & $-0.490~(0.266)~$  & $-0.488~(0.566)~$   & $1.004~(0.470)$   \\
\texttt{blossom:des}        & $-1.235~(0.281)^*$ & $-1.242~(0.604)^*$  & $0.994~(0.465)$   \\
\texttt{blossom:des$^2$}    & $0.665~(0.316)^*$  & $0.673~(0.680)~$    & $0.989~(0.465)$   \\
\texttt{fig:des}            & $0.380~(0.265)~$   & $0.365~(0.566)~$    & $1.040~(0.468)$   \\
\texttt{fig:des$^2$}        & $-1.330~(0.313)^*$ & $-1.310~(0.673)~$   & $1.015~(0.465)$   \\
\texttt{boll:des}           & $0.011~(0.237)~$   & $0.009~(0.504)~$    & $1.181~(0.471)$  \\
\texttt{boll:des$^2$}       & $-0.021~(0.260)~$  & $-0.020~(0.553)~$   & $1.059~(0.471)$   \\ \hline
$p$                         & $0.981~(0.137)~$   & $-$                 & $-$   \\
$\phi$                      & $-0.810~(0.223)~$  & $-$                 & $-$    \\ \hline     
\end{tabular}
\end{table}

\begin{figure}[h]
\centering
\includegraphics{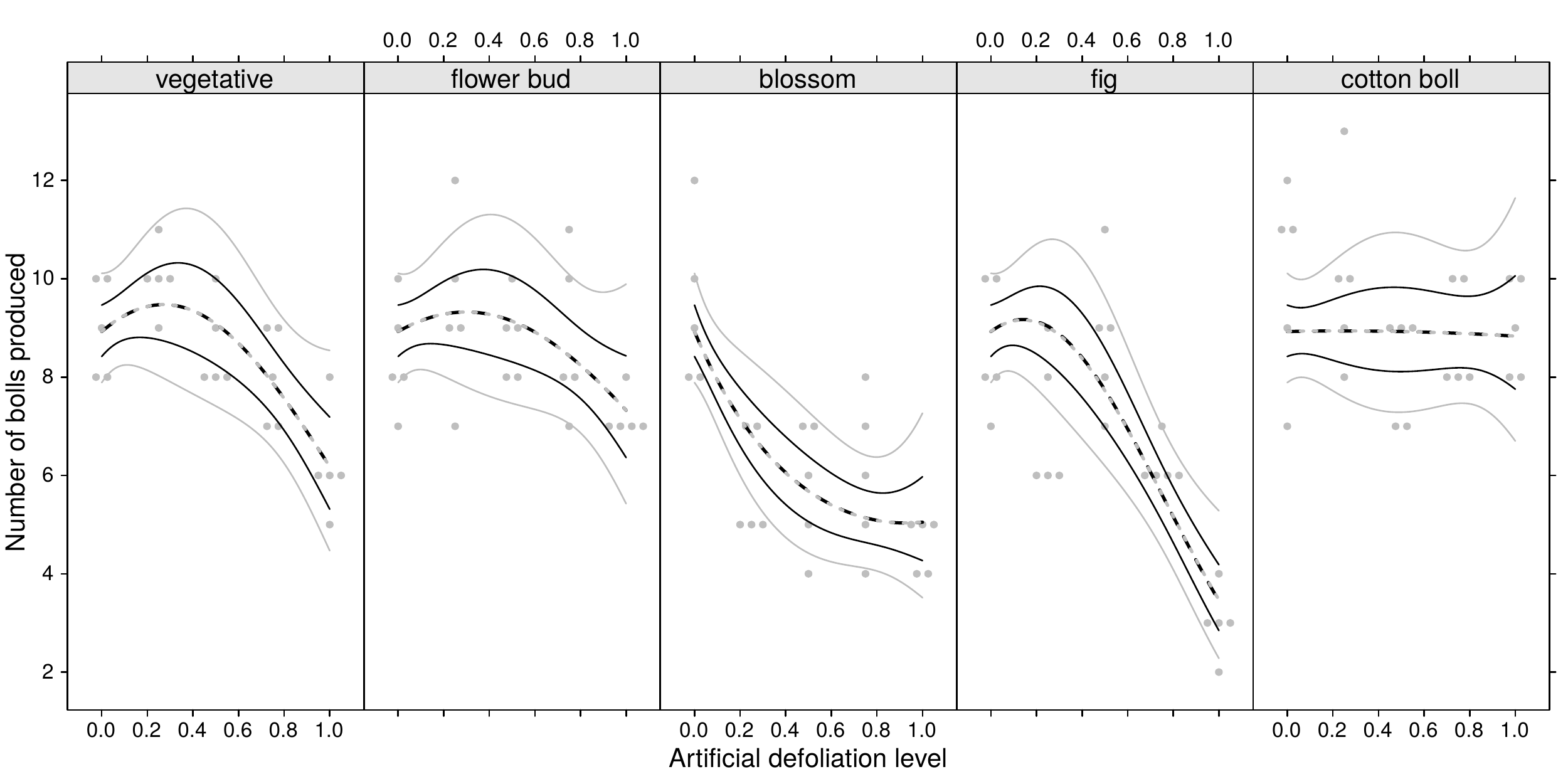}
\caption{Dispersion diagrams of observed values and curves of fitted 
values~(Poisson-gray and Poisson-Tweedie-black) and confidence intervals $(95\%)$ 
as functions of the defoliation level for each growth stage.}
\label{fig:example2}
\end{figure}

The results in Table~\ref{tab:example2} show that the estimates are 
quite similar, however, the standard errors obtained by the 
Poisson-Tweedie model are smaller than those from the Poisson model. 
This is explained by the negative estimate of the dispersion parameter, 
which indicates underdispersion. The value of the power parameter is
close to $1$ and explains the similarity of the regression parameter estimates. 
Appropriate estimation of the standard error is important for this data set, 
since the Poisson-Tweedie identifies the effect of the 
defoliation as significant for three of the five growth stages, while the Poisson model only finds 
the defoliation effect as significant for the blossom growth stage. 
Figure~\ref{fig:example2} presents the observed values and curves of 
fitted values~(Poisson in gray and Poisson-Tweedie in black) and 
confidence intervals ($95\%$) as functions of the 
defoliation level for each growth stage and supports the above conclusions.

\setkeys{Gin}{width=0.99\textwidth}

The results from the Poisson-Tweedie model are consistent with those from
the Gamma-Count model, fitted by~\citet{Zeviani:2014},
in that both methods indicate underdispersion and significant effects of 
defoliation for the vegetative, blossom and fig growth stages .
However, it is important to note that the estimates obtained by the Gamma-Count 
model fitted by~\citet{Zeviani:2014} are not directly comparable with 
the ones obtained from the Poisson-Tweedie model, since the latter is 
modelling the expectation, while the Gamma-Count distribution  
models the distribution of the time between events.

\subsection{Data set 3: radiation-induced chromosome aberration counts}
In this example, we apply the extended Poisson-Tweedie model to describe
the number of chromosome aberrations in biological dosimetry.
The dataset considered was obtained after irradiating blood samples
with five different doses between $0.1$ and $1$ Gy of $2.1$ MeV neutrons. 
In this case, the frequencies of dicentrics and centric rings after a culture of $72$ hours are analysed.
The dataset in Table~\ref{tab:data3} was first presented by~\citet{Heimers:2006}
and analysed by~\citet{Oliveira:2015} as an example of zero-inflated data.

\begin{table}[h]
\centering
\caption{Frequency distributions of the number of dicentrics and centric 
rings by dose levels.}
\label{tab:data3}
\begin{tabular}{lrrrrrrrr} \hline
\multirow{2}{*}{$x_i$}  & \multicolumn{8}{l}{ $y_{ij}$ } \\ \cline{2-9}
      & $0$    & $1$    & $2$   & $3$  & $4$  & $5$  & $6$  & $7$ \\ \hline
$0.1$ & $2281$ & $130$  & $21$  & $1$  & $0$  & $0$  & $0$  & $0$    \\
$0.3$ & $847$  & $127$  & $19$  & $6$  & $1$  & $0$  & $0$  & $0$    \\
$0.5$ & $567$  & $165$  & $49$  & $16$ & $2$  & $0$  & $0$  & $0$  \\
$0.7$ & $356$  & $167$  & $62$  & $9$  & $5$  & $1$  & $0$  & $0$  \\
$1$   & $169$  & $131$  & $72$  & $18$ & $9$  & $0$  & $0$  & $1$  \\ \hline
\end{tabular}
\end{table}

We fitted the extended Poisson-Tweedie and Poisson models with the 
linear predictor specified as a quadratic dose model, i.e
$$ g(\mu_{ij}) = \beta_0 + \beta_{1} \texttt{dose}_i + \beta_{2} \texttt{dose}^2_i.$$ 
Table~\ref{tab:example3} presents the estimates and standard errors for 
the Poisson-Tweedie and Poisson models, along with the ratios between the respective
estimates and standard errors.

\begin{table}[h]
\centering
\caption{Data set $3$: Parameter estimates and standard errors (SE) for Poisson-Tweedie 
and Poisson models (first and second columns). Ratios between 
Poisson-Tweedie and Poisson estimates and standard errors (third column).}
\label{tab:example3}
\begin{tabular}{lrrrr} \hline
\multirow{2}{*}{Parameter}     &\multicolumn{3}{c}{Estimates (SE)}   \\ \cline{2-4} 
                     & Poisson-Tweedie    & Poisson               & Ratio  \\ \hline
$\texttt{Intercept}$ & $-3.126~(0.106)^*$ & $-3.125~(0.097)^*$ & $1.000~(1.098)$  \\
$\texttt{dose}$      & $5.514~(0.408)^*$  & $5.508~(0.369)^*$  & $1.001~(1.104)$   \\
$\texttt{dose}^2$    & $-2.481~(0.342)^*$ & $-2.476~(0.309)^*$ & $1.002~(1.107)$   \\
$p$                  & $1.085~(0.299)~$   & $-$                   & $-$   \\
$\phi$               & $0.249~(0.100)~$   & $-$                   & $-$    \\ \hline     
\end{tabular}
\end{table}

Results in Table~\ref{tab:example3} show evidence of weak overdispersion that can 
be attributed to zero-inflation, since the estimate of the power 
parameter was close to $1$, which in turn implies that the
standard errors obtained from the Poisson-Tweedie model are around $10\%$ 
larger than those obtained from the Poisson model. 

For this data set it is particularly easy to compute the log-likelihood value, 
since we have only a few unique observed counts and dose values. 
Thus, we can use log-likelihood values to compare the fit of the Poisson-Tweedie model with the fit obtained 
by the zero-inflated Poisson and zero-inflated negative binomial models. The log-likelihood value of the 
Poisson-Tweedie model was $-2950.605$, while the maximised log-likelihood 
value of the zero-inflated Poisson and zero-inflated negative binomial
models were $-2950.462$ and $-2950.531$, respectively. 
Furthermore, the maximised log-likelihood value of the Poisson model was 
$-2995.389$. These results show that the Poisson-Tweedie model can offer a 
very competitive fit, even without an additional linear predictor to describe 
the excess of zeroes. Furthermore, it is interesting to note that in 
spite of the large difference in  the log-likelihood values, 
the Poisson model provides the same interpretation in terms of the 
significance of the covariates as the Poisson-Tweedie model 
for this data set.

\subsection{Data set 4: customers' profile}
The last example corresponds to a data set collected to investigate the 
customer profile of a large company of household supplies. 
During a representative two-week period, in-store surveys were conducted 
and addresses of customers were obtained. The addresses were then used 
to identify the metropolitan area census tracts in which the customers 
resident. At the end of the survey period, the total number of customers 
who visited the store from each census tract within a 10-mile radius was 
determined and relevant demographic information for each tract was obtained. 
The data set was analysed in~\citet{Neter:1996} as an example of Poisson 
regression model, since it is a classic example of equidispersed count data.
Following~\citet{Neter:1996} we considered the covariates, 
number of housing units~(\texttt{nhu}), average income in dollars~(\texttt{aid}), 
average housing unit age in years~(\texttt{aha}), distance to the nearest 
competitor in miles~(\texttt{dnc}) and distance to store in miles~(\texttt{ds}) 
for forming the linear predictor.

For equidispersed data the estimation of the Tweedie power parameter is in 
general a difficult task. In this case, the dispersion parameter $\phi$ 
should be estimated around zero. Thus, we do not have enough information 
to distinguish between different values of the Tweedie power parameter. 
Consequently, we can fix the Tweedie power parameter 
at any value and the corresponding fitted models should be very similar. 
To illustrate this idea, we fitted the extended Poisson-Tweedie model
fixing the Tweedie power parameter at the values $1$, $2$ and $3$, corresponding
to the Neyman Type A (NTA), negative binomial (NB) and Poisson-inverse 
Gaussian (PIG) distributions, respectively. 
We also fitted the standard Poisson model for comparison, the estimates and 
standard errors (SE) are presented in Table~\ref{tab:example4}.

\begin{table}[h]
\centering
\caption{Data set $4$: Estimates and standard errors (SE) from different models.}
\label{tab:example4}
\begin{tabular}{lrrrr}\hline
Parameter            & Poisson            & NTA                 & NB                 & PIG          \\ \hline
\texttt{Intercept}   & $2.942~(0.207)^* $ & $2.942~(0.194)^*$  & $2.937~(0.197)^*$  & $2.933~(0.203)^*$ \\ 
\texttt{nhu}         & $0.061~(0.014)^* $ & $0.061~(0.013)^*$  & $0.060~(0.013)^*$  & $0.060~(0.014)^*$\\ 
\texttt{aid}         & $-0.012~(0.002)^*$ & $-0.012~(0.002)^*$ & $-0.012~(0.002)^*$ & $-0.012~(0.002)^*$ \\ 
\texttt{aha}         & $-0.004~(0.002)^*$ & $-0.004~(0.002)^*$ & $-0.004~(0.002)^*$ & $-0.004~(0.002)^*$ \\ 
\texttt{dnc}         & $0.168~(0.026)^*$  & $0.168~(0.024)^*$  & $0.165~(0.025)^*$  & $0.166~(0.025)^*$ \\ 
\texttt{ds}          & $-0.129~(0.016)^*$ & $-0.129~(0.015)^*$ & $-0.127~(0.015)^*$ & $-0.127~(0.016)^*$ \\ \hline
$\phi$               & $0$                & $-0.122(0.123)~$   & $-0.008~(0.010)$   & $0.000~(0.000)$ \\
$p$                  & $-$                & $1$                & $2$                & $3$           \\ \hline
\end{tabular}
\end{table}

The results presented in Table~\ref{tab:example4} show clearly that for all fitted models the dispersion
parameter does not differ from zero, which gives evidence of equidispersion. The regression coefficients and the
associated standard errors do not depend on the models and in particular do not depend on the power 
parameter value. This example shows that, although a more careful analysis is required, the extended 
Poisson-Tweedie model can deal with equidispersed data.
Furthermore, the estimation of the extra dispersion parameter does not inflate the standard errors
associated with the regression coefficients. Thus, there is no loss of efficiency when using the
Poisson-Tweedie model for equidispersed count data.

\section{Discussion}~\label{discussion}

We presented a flexible statistical modelling framework to deal with count data.
The models are based on the Poisson-Tweedie family of distributions that automatically
adapts to overdispersed, zero-inflated and heavy-tailed count data.
Furthermore, we adopted an estimating function approach for estimation and
inference based only on second-order moment assumptions. Such a specification 
allows us to extend the Poisson-Tweedie model to deal with underdispersed
count data by allowing negative values for the dispersion parameter.
The main technical advantage of the second-order moment specification is
the simplicity of the fitting algorithm, which amounts to finding the root
of a set of non-linear equations. The Poisson-Tweedie family encompasses
some of the most popular models for count data, such as the Hermite,
Neyman Type A, P\'olya-Aeppli, negative binomial and Poisson-inverse
Gaussian distributions. For this reason, the estimation of the power parameter
plays an important role in the context of Poisson-Tweedie regression models,
since it is an index that distinguishes between these important distributions.
Thus, the estimation of the power parameter can work as an automatic 
distribution selection.

We conducted a simulation study on the properties of the estimating
function estimators. The results showed that in general the estimating function 
estimators are unbiased and consistent. 
We also evaluated the validity of the standard errors obtained by the estimating function approach 
by computing the empirical coverage rate. The results showed that for the regression 
coefficients our estimators provide the specified level of coverage for all 
simulation scenarios and sample sizes. Regarding the dispersion parameter, the results
showed that for small samples the standard errors are underestimated, however,
the results improve for larger samples. On the other hand, the standard errors associated
with the power parameter are overestimated for all simulation scenarios and sample sizes.
However, the coverage rate presented values only slightly larger than the specified nominal
level of $95\%$. It is important to highlight that the under or overestimation of the
dispersion and power parameters do not affect the estimates and standard errors associated
with the regression coefficients. This is due to the insensitivity property, 
see equation~(\ref{Sbetalambda}). Furthermore, we demonstrated the flexibility of the extended 
Poisson-Tweedie model to deal with underdispersed count data as generated by the COM-Poisson
and Gamma-Count distribution. It also shows that the model has a good level of robustness
against model misspecification.

Discussion of the efficiency of the estimating function estimators is difficult due 
to the lack of a closed form for the Fisher information matrix.
\citet{Bonat:2016a} showed in the context of Tweedie regression models that the quasi-score
function provides asymptotically efficient estimators for the regression parameters, thus
a similar result is expected for the Poisson-Tweedie regression model. Concerning the dispersion
and power parameters, the fact that the sensitivity and variability matrices do not coincide
indicates that the Pearson estimating functions are not optimum. Furthermore, the use of 
empirical third and fourth moments for the calculation of the Godambe information matrix
must imply some efficiency loss. On the other hand, it again makes the model robust against
misspecification. 

We analysed four real data sets to explore and illustrate the flexibility of the extended 
Poisson-Tweedie model. Data set $1$ presented a classical case of overdispersion.
This data set illustrated the most common problem when using the Poisson model for overdispersed 
count data, i.e. the strong underestimation of the standard errors associated with the
regression coefficients. The Poisson-Tweedie model automatically adapts to the dispersion
in the data by the estimation of the dispersion parameter, while choosing the appropriate distribution
in the Poisson-Tweedie family through the estimation of the power parameter. Furthermore, 
the uncertainty around the data distribution is taken into account and can be assessed based
on the standard errors associated with the power parameter. In particular, for this application
the model shows that any distribution in the family of the Poisson compound Poisson distributions
$(1 < p < 2)$ provides a suitable fit for the data set. 
Thus, we avoid the need to fit an array of models and the use of measures of goodness-of-fit
to choose between them.

Data set $2$ presents the less frequent case of underdispersion. In this case, the problem
is that the Poisson model overestimates the standard errors associated with the regression 
coefficients. The negative value of the dispersion parameter obtained by fitting the Poisson-Tweedie
model to this data set indicates underdispersion. Thus, the model automatically corrects the 
standard errors for the regression coefficients, giving standard errors that are smaller 
than those obtained from the Poisson model. The problem of zero-inflated count data was 
illustrated by the data set $3$. In this example, we showed that, in general, zero-inflation
introduces overdispersion and that the Poisson-Tweedie model can also adapt to zero-inflation
providing a very competitive fit when compared with more orthodox approaches such as the 
zero-inflated Poisson and zero-inflated negative binomial models. Finally, data set $4$
illustrated the case of equidispersed count data. This case is particularly challenging for the
Poisson-Tweedie model since the dispersion parameter should be zero, which implies that 
any distribution in the family of Poisson-Tweedie distributions can provide a suitable fit
for the data. Thus, the estimation of the Tweedie power parameter is very difficult, because the
estimating function associated with the Tweedie power parameter is flat. In this case, our approach
was to fit the model with the Tweedie power parameter fixed at the values $1$, $2$ and $3$.
We compared the fit of these three models with the fit of the Poisson model and, since we have
equidispersed data, all models provide quite similar estimates and standard errors.
Furthermore, all models indicated that the dispersion parameter is not different from
zero, which again indicates equidispersion. It is important to emphasize that the estimation
of the additional dispersion parameter does not inflate the standard errors associated with
the regression parameters.

There are many possible extensions to the basic model discussed in the present paper, 
including incorporating penalized splines and the use of regularization for high dimensional 
data, with important applications in genetics. There is also a need to develop methods for model
checking, such as residual analysis, leverage and outlier detection.
Finally, we can extend the model to deal with multivariate count data, with
many potential applications for the analysis of longitudinal and spatial data.
These extensions will form the basis of future work.

\section*{Acknowledgements}
This paper is dedicated in honour and memory of Professor Bent J{\o}rgensen.
This work was done while the first author was visiting the Laboratory of Mathematics 
of Besan\c{c}on, France and School of Mathematics of National University 
of Ireland, Galway, Ireland.
The first author is supported by CAPES (Coordena\c{c}\~ao de
Aperfei\c{c}omento de Pessoal de N\'ivel Superior), Brazil.
The last author was partially supported by CNPq, a Brazilian Science Funding Agency.

\bibliographystyle{dcu} 
\bibliography{Bonatetal2016A}

\end{document}